\documentclass[aip,jcp,reprint,longbibliography,nofootinbib]{revtex4-2}

\usepackage{amsmath,amssymb,mathtools,bm}
\usepackage{graphicx}
\usepackage{booktabs}
\usepackage{array}
\usepackage{placeins}
\usepackage{microtype}
\usepackage[dvipsnames]{xcolor}
\usepackage[colorlinks=true,linkcolor=MidnightBlue,citecolor=MidnightBlue,urlcolor=MidnightBlue]{hyperref}

\graphicspath{{figs/}{./}}

\newcommand{\dd}{\mathrm d}

\newcommand{\Span}{\mathrm{span}}
\newcommand{\PQS}{\mathrm{PQS}}

\begin{document}

\title{Projected \(q\)-Shells for Nested Gausslet Bases}

\author{Steven R. White}
\affiliation{Department of Physics and Astronomy, University of California, Irvine, Irvine, CA 92697, USA}

\date{August 27, 2026}

\begin{abstract}
Electronic-structure calculations require a finite representation of both the
electronic wave functions and their Coulomb interaction.  Conventional compact
basis sets lead to a four-index interaction. Gausslet basis sets are one of a
very small number of approaches that allow a much smaller two-index
interaction, similar to a real-space grid, but with far fewer points. The
continuing challenge in their development is reducing further the number of
functions needed while maintaining high accuracy. Here we introduce projected
\(q\)-shells (PQS), an improvement to the nested Cartesian gausslets of White
and Lindsey. Their standard nesting construction divides a joining shell into
disjoint faces, edges, and corners, reducing its polynomial completeness from
order \(q\) to order \(q-2\). PQS approximately restores the two lost orders.
In terms of completeness, we prove that for an ideal, undistorted parent,
the span of the PQS basis is equivalent to an overcomplete basis of overlapping patches.
Coordinate distortions, which are always used in practice, harm this equivalence only slightly.
The moment properties that allow two-index interactions also become inexact, even without
distortion, but we show that the improvement in order more than makes up for this loss.
For similar basis sizes, PQS gives smaller
one-electron and interaction errors on small molecules, such as H$_2^+$, H$_2$,
and C$_2$, than standard nesting. The
improvement is largest for more compact bases with smaller $q$.
\end{abstract}

\maketitle

\section{Introduction and background}
\label{sec:intro}

Electronic-structure calculations for atoms, molecules, and solids begin by
replacing the continuous spatial dependence of the electrons by a finite
representation.  The choice of that representation strongly affects both the
number of functions needed for a given accuracy and the cost of treating the
electron--electron interaction.  Atom-centered Gaussian orbitals can give a
compact description of molecular wave functions, but their Coulomb integrals
form a four-index tensor.  A real-space grid has the opposite advantage: local
potentials and the Coulomb interaction have a much simpler form, but a uniform
grid needs many points to resolve the very different length scales near and far
from a nucleus.  

Gausslets were introduced to have the diagonal two-index interaction of a grid while
having far fewer grid points
\cite{White2017Gausslets,WhiteStoudenmire2019}.  They are smooth, localized,
orthonormal functions that act approximately like delta functions 
when integrated against smooth functions.  This last property permits the
four-index electron--electron interaction to be replaced accurately by a
two-index density--density interaction.  The reduction is especially useful
in correlated calculations, where storing and repeatedly applying the
interaction can otherwise be a dominant cost.  Cartesian gausslets, hybrid
gausslet--Gaussian bases, and nested gausslets have been tested on atoms, small
molecules, and hydrogen chains, including calculations with the density matrix
renormalization group
\cite{WhiteStoudenmire2019,QiuWhite2021,WhiteLindsey2023}.  Gaussian basis supplements
to a main gausslet basis~\cite{QiuWhite2021}
are particularly useful for the sharp core and cusp structure that would be
expensive to describe by local refinement alone.

The main limitation of current gausslet constructions is geometric.  The
basic functions are known in one dimension, and three-dimensional bases have
therefore usually been formed as Cartesian products.  In one dimension a
coordinate transformation can place many functions near a nucleus and fewer
far away.  Applying three such transformations independently, however, refines
entire Cartesian slabs rather than a compact neighborhood of the nucleus.
Radial and angular gausslets provide a natural atom-centered alternative
\cite{White2026RadialGausslets,White2026AngularGausslets}, but Cartesian product
bases remain more general for molecules and for regions not naturally assigned
to a single center.

The White--Lindsey (WL) nested gausslet construction addresses this geometric problem by contracting a
parent Cartesian product basis into a sequence of nested shells
\cite{WhiteLindsey2023}.  The shells contain the same number of functions even
as their physical scale changes by a large factor, so the finest Cartesian
grid need not extend through the whole molecule.  The construction has a
weakness, however: each shell is chopped into disjoint flat pieces---faces,
edges, and corners.  These smaller pieces reduce the completeness order by
two.  Whereas a filled contracted three-dimensional cube of size \(q\)
represents all polynomials through degree \(q-1\), the disjoint pieces are
complete only through degree \(q-3\).  Counting the constant function, we
refer to these as order \(q\) and order \(q-2\), respectively.  The loss is
modest at high local order, but it can be important in the compact bases needed to
do larger systems.
For \(q=5\), for example, it is the difference between retaining quartic and
quadratic variation along a contracted direction.

The aim of this work is to retain the full local order without increasing the
nominal shell dimension.  The PQS approach is based on a surprising result,
the projected-shell theorem, proved in Appendix~\ref{app:patchproof}.  It shows
that a manifestly complete but highly redundant set of overlapping patches is
equivalent in span to a much smaller nested shell construction.  The disjoint
flat pieces are avoided in PQS and there is no loss of completeness order in the
ideal construction.

Two qualifications arise when this idea is used for molecular gausslets.
Even without coordinate distortion, projection onto the shell makes the
moment properties of the individual functions inexact, although it preserves
the span of the complete patches.  In addition, the standard one-dimensional
coordinate mapping makes that span equivalence only approximate.  We measure
both effects directly before turning to electronic-structure tests.  The
results show that
the advantage of retaining the full patch order more than compensates for
these departures from the ideal construction: PQS performs significantly
better than standard nesting
both as a one-particle basis and when used with the diagonal interaction.

Section~\ref{sec:pqs} develops the construction and tests its span, moments,
and representation of smooth functions.  Section~\ref{sec:evidence} then
compares PQS directly with WL nesting for H$_2^+$, H$_2$, and
restricted C$_2$, testing one-particle contraction, molecular energy curves,
and multi-orbital states at matched or nearly matched dimensions.  The proof
of the projected-shell theorem is given in
Appendix~\ref{app:patchproof}.

\section{Projected \(q\)-shell nested gausslet bases}
\label{sec:pqs}

\begin{figure}[t]
\centering
\includegraphics[width=0.85\columnwidth]{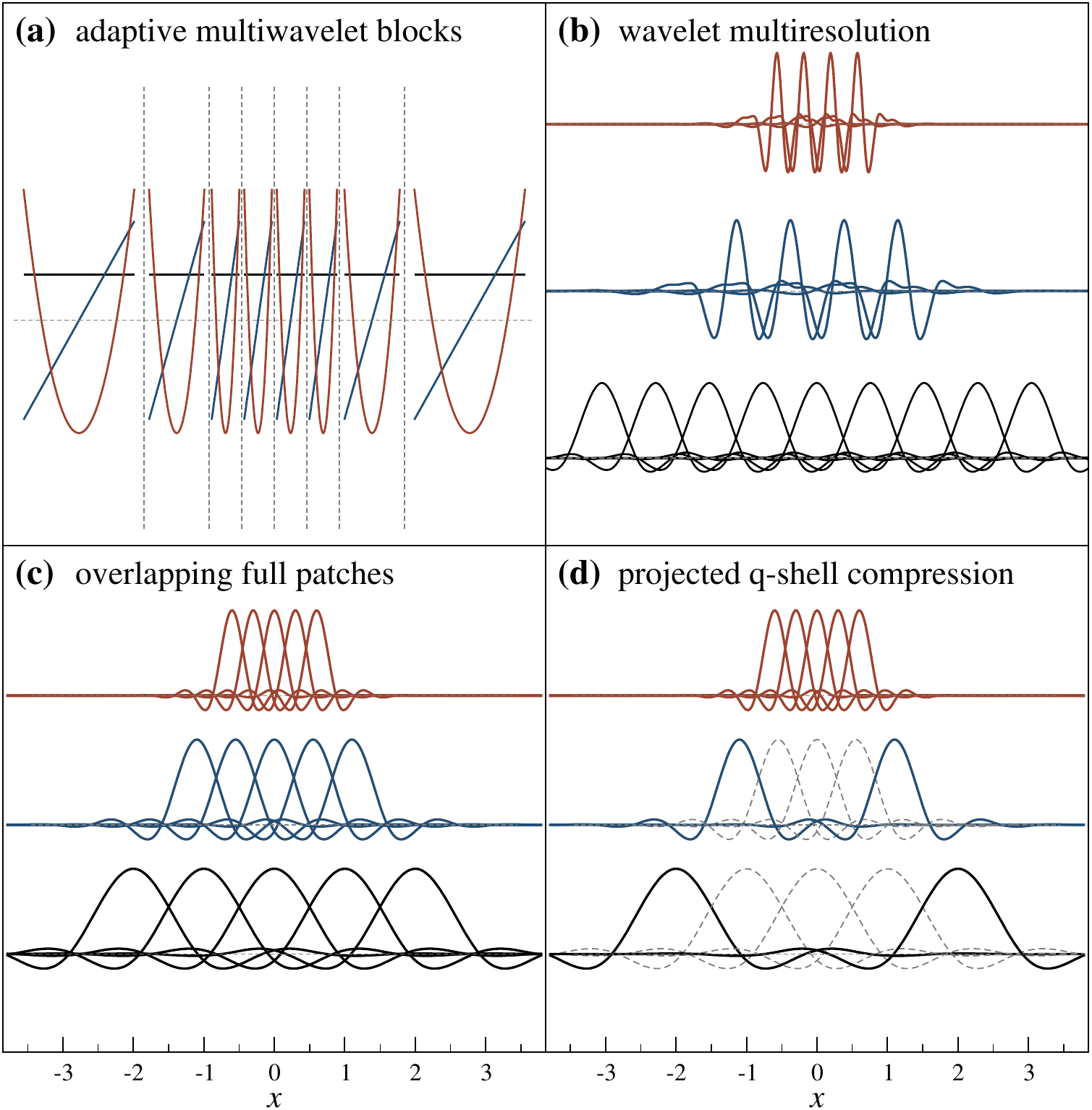}
\caption{Four ways of changing spatial resolution.  (a) Adaptive
multiwavelet blocks.  (b) Wavelet multiresolution.  (c) Overlapping full
gausslet patches at several scales.  (d) Projected \(q\)-shell
compression of the same patches.  The dashed interior functions in (d) are
removed because their space is already supplied by the finer patch.}
\label{fig:multiresolution-patches}
\end{figure}

The purpose of nesting is to use a fine basis only where it is needed, while
retaining smooth localized functions with zero moments throughout.  Seams where the
resolution changes abruptly almost always destroy the moment properties and require
careful design to avoid loss of completeness.
Figure~\ref{fig:multiresolution-patches}
illustrates this problem and the construction used here.

The figure first shows two familiar approaches. Adaptive multiwavelets, panel (a), 
subdivide selected cells and retain a small
polynomial basis in each cell \cite{Harrison2004,Fann2004Singular}.  They give
systematic local refinement and strict compact support, but the cell boundary
allows discontinuities which must be controlled, and each cell retains a small four-index interaction.
Ordinary wavelet multiscale decompositions, panel (b), instead give an exact hierarchy of coarse scaling
functions and fine detail functions \cite{DaubechiesBook}.  Their multiscale oscillatory functions
are very useful for compression, but they do not allow a diagonal interaction. For a diagonal approximation
to work, each function should be sharply peaked at a separate point in space, 
to act like a broadened delta function there.

The third panel shows another way to get completeness at a nucleus: combine patches of gausslets at
different scales, all centered on the nucleus.  These are contractions of nested regions of the same
parent lattice, rather than functions constructed on separate grids. It is easy to see that the completeness of such a 
construction is excellent: each length scale has its own set of functions that resolve it. The problem is that
it is not only nonorthogonal, it is overcomplete, and its overlap matrix would have many near-zero eigenvalues.
In addition, the functions are not peaked at separate locations, so a diagonal interaction is not possible.

The final panel shows the projected \(q\)-shell (PQS) construction,
schematically in 1D. Here, the interiors of the patches in (c) are deleted.
Remarkably, a new theorem, which we call the projected-shell theorem, shows
that with the right
construction of the functions, the span of (d) is equal to that of (c) in the ideal uniform construction, and the extra functions in (c) have
exactly zero eigenvalues in the overlap matrix. The PQS construction retains the resulting high completeness while
keeping the moment errors sufficiently small for diagonal interactions.

\subsection{Gausslets, COMX, and nesting}

Here we give some minimal background on gausslet bases. A more complete introduction is given 
in \cite{WhiteLindsey2023}.
A gausslet \(G_i(\mathbf r)\) is localized about a center \(\mathbf r_i\)
and integrates smooth functions approximately as a weighted delta function.
For the electron--electron interaction this permits the four-index Coulomb
tensor to be replaced by a two-index diagonal form,
\begin{equation}
V_{ijkl}\simeq \delta_{ij}\delta_{kl}V_{ik}.
\label{eq:gausslet-diagonal}
\end{equation}
The approximation is not based merely on small off-diagonal integrals.  It
reproduces the action of the complete interaction on smooth states, and its
accuracy depends on the combination of locality, orthogonality, polynomial
completeness, and moment properties of the basis.

The COMX theorem~\cite{WhiteLindsey2023}, a key ingredient of nested gausslets,
links completeness (C), orthonormality (O), the gausslet moment conditions (M),
and diagonalization of the coordinate matrix (X).  For a 1D line of functions,
completeness, orthonormality, and coordinate-matrix diagonalization together
produce the moment conditions needed for the diagonal interaction.  In the
nesting construction, these 1D COMX contractions are used either in 1D form
for shell edges or in tensor-product form for 2D faces.  Complete 3D blocks
use a third tensor-product direction.  Gausslets themselves already possess
all four properties to high order.  In nesting, COMX contracts a large,
distorted parent gausslet basis into a much smaller basis.  A 1D contraction
fits a line of parent functions to a smaller set of low-order polynomials,
then orthonormalizes and coordinate-diagonalizes the resulting functions.
Its completeness order is controlled by the number of retained functions,
\(q\).

In one dimension, distortion is controlled by a monotone coordinate map \(t(x)\) via
\begin{equation}
\widetilde G_i(x)=\sqrt{t'(x)}\,G\!\left(t(x)-i\right).
\label{eq:distorted-gausslet}
\end{equation}
The transformation preserves orthonormality and, when the spacing changes
slowly, retains the moment accuracy needed for the diagonal approximation
\cite{WhiteStoudenmire2019,WhiteLindsey2023}.  All scales are contractions of
nested regions of this one mapped parent lattice, not functions built on
separate grids.  Larger source regions, together with the increasing physical
spacing of the mapped lattice, produce the broader outer functions.  Three independent Cartesian
maps, however, refine whole slabs whenever one coordinate lies near a nucleus.
Nested gausslets avoid this cost by keeping a fine central block and
surrounding it by progressively coarser shells.

In WL nesting, the COMX contractions can only be done on disjoint sets of functions. They are normally
done in shells; consider a $q^3$ cube of functions. The outer layer, or shell of the cube is broken into disjoint
pieces, resulting in faces of linear size $q-2$, as well as edges and corners. This makes the order of the nesting
$q-2$, a sizable loss, for, say, $q=5$.
PQS is designed to recover this lost local order without changing the cube size.

\begin{figure}[t]
\centering
\includegraphics[width=0.85\columnwidth]{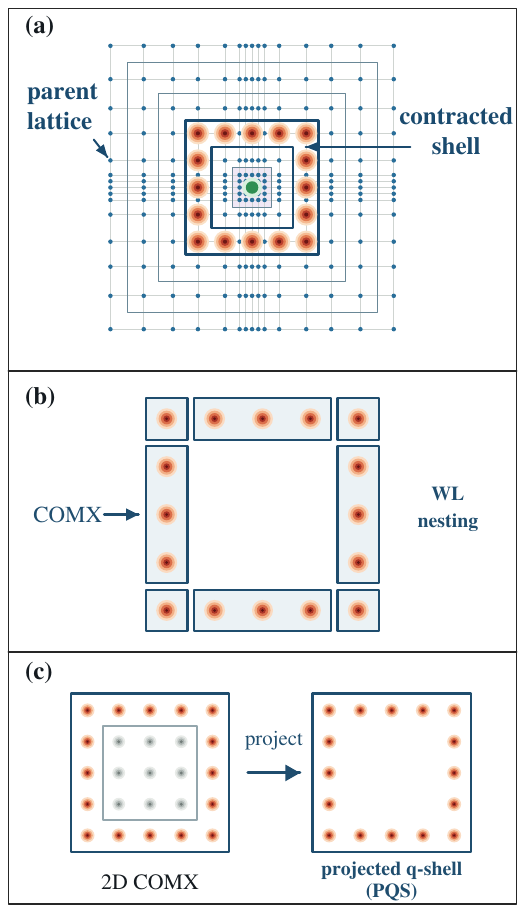}
\caption{\label{fig:pqs-2d-shell}Parent lattice and two-dimensional \(q=5\)
shell.  (a) A distorted parent lattice, the uncontracted central block, and
one surrounding shell.  (b) WL treats the corners and the four
three-function sides separately.  (c) PQS starts from the complete
\(5\times5\) local block.  Its 16 products having at least one edge factor are
projected onto the shell of parent-lattice sites shown in (a) and then
reorthogonalized; the nine middle--middle products are omitted.  Panels (b)
and (c) have the same number of functions but retain different local
polynomial information.}
\end{figure}

\subsection{Projected \(q\)-shell construction}

The PQS construction in 2D is shown in Fig.~\ref{fig:pqs-2d-shell}.  The
parent lattice shown in panel (a) contains a set of concentric, overlapping
squares of parent functions surrounding a nucleus.  Each square is filled:
all parent functions inside it are included.  As shown in panel (c), the
parent functions in one such square are contracted using a pair of 1D COMX
contractions into an array of \(q^2\) functions.  The outer shell of
contracted functions is then projected onto the corresponding one-site-thick
shell of parent-lattice sites.
The projection destroys the orthonormality of the 2D COMX functions, so we
symmetrically orthonormalize the functions in the shell.  These three
operations---the full COMX contraction, projection, and
reorthogonalization---make up the basic PQS operation.  The smallest
\(q^2\) square of functions at the nucleus is retained in full after
contraction.  In contrast, panel (b) shows the simpler WL nesting,
where the shell is broken into disjoint flat pieces.

In one dimension, the square is a line segment, and the shell consists only
of its two endpoint sites.  A COMX is performed on the whole line segment,
producing \(q\) COMX functions, of which the two outer, or edge, functions are
selected as candidates for the new shell.  Each candidate still has
coefficients on parent gausslets throughout its source segment.  We project
them onto the corresponding shell of parent-lattice sites.  In 1D the
projected pair spans the two endpoint parent functions; with reflection
symmetry, symmetric orthogonalization gives the endpoint parent functions
themselves.  The process appears trivial, but its span identity is
not.  The inner line of parent functions---excluding the two endpoints---is
separately contracted into \(q\) COMX functions.  The projected-shell theorem
proved in Appendix~\ref{app:patchproof} shows that these \(q+2\) retained
functions have the same span as the full set of \(2q\) inner- and outer-line
COMX functions.  Even more importantly, the theorem applies in higher
dimensions, where projection onto the shell is not trivial.

Figure~\ref{fig:one-dimensional-projected-shell}(b) in
Appendix~\ref{app:patchproof} shows directly that 1D projection onto the
parent shell changes the two edge functions only modestly.  As developed in
the Appendix, the COMX centers, denoted there by \(\lambda_a\), are the
Gaussian-quadrature nodes associated with the parent sites.  Like ordinary
Gaussian quadrature points, they crowd toward the two ends of an interval.
The two outer COMX functions are therefore already concentrated near the
endpoint parent sites, with relatively little weight on the parent sites
removed by the projection.  Endpoint crowding becomes stronger as \(q\)
grows, so projection changes these functions less at higher order.  Moment
accuracy is a separate
issue from completeness.  Here completeness refers primarily to the
polynomial order of the retained space, together with its ability to resolve
smooth functions, whereas the moment conditions are properties of each
individual function.  For the ideal COMX functions the moment conditions are
exact before this projection.  Projection makes them inexact.  Its modest
effect on the functions suggests that the resulting moment errors should be
small, but this must be tested; we do so in
Sec.~\ref{subsec:pqs-construction-tests}.

In 3D the construction is applied to cubes or rectangular boxes.  The box
COMX is a tensor product of three 1D COMX contractions.  The outer shell
consists of every product having at least one of the two edge COMX functions
along one axis.  These functions are projected onto the three-dimensional
shell of parent sites of that box.  For a diatomic molecule, a block
containing both nuclei is generally rectangular, \(q\times q\times L\), with
its long direction along the bond.  There are \((q-2)^2(L-2)\) all-middle
products, leaving a projected-shell dimension
\begin{equation}
N_{\PQS}=q^2L-(q-2)^2(L-2).
\label{eq:pqsdim}
\end{equation}
For a cubic \(q=5\) patch this leaves \(125-27=98\) functions, exactly the
same number as in WL nesting.  PQS starts with a complete \(q=5\) contraction
and makes its moment properties approximate through projection and
reorthogonalization, whereas WL builds its faces, edges, and corners as
separate exact lower-order pieces.

More explicitly, the PQS construction is defined in terms of matrices acting
on the parent functions.  Let the columns of \(C\) be the 3D full
COMX contractions defining functions on the 3D parent block.  For simplicity,
take \(L=q\), and let the parent cube be of size \(k\).  Then
\(C\) is \(k^3\times q^3\).  Let \(P\) select the shell
of COMX functions, with dimensions
\(q^3\times[q^3-(q-2)^3]\), and let \(Q\), with dimensions
\([k^3-(k-2)^3]\times k^3\), restrict the parent coefficients to the
one-site-thick parent shell.  The raw shell columns and their symmetrically
orthonormalized form are
\begin{equation}
Y=QCP,
\qquad
B=Y(Y^\dagger Y)^{-1/2}.
\label{eq:lowdin}
\end{equation}
The selected columns of \(C\) are mutually orthonormal before the
shell projection.  Projecting them onto the parent shell generally destroys
both their normalization and their mutual orthogonality.  The second
operation in Eq.~\eqref{eq:lowdin} restores orthonormality without changing
their shell-supported span, giving \(B^\dagger B=I\).

On an ideal uniform parent, the inner contracted cube together with the
projected outer shell spans the same space as the complete inner and outer
contracted cubes.  Shell projection and reorthogonalization nevertheless make
the individual moment conditions inexact even without coordinate distortion.
Coordinate mapping introduces a separate error: it breaks the exact span
identity and can therefore weaken completeness.  We test the two effects
separately below.

\subsection{Tests of the construction}
\label{subsec:pqs-construction-tests}

We begin with direct numerical tests of the two properties on which PQS
relies: the span identity that preserves completeness and the moment
conditions needed for the diagonal interaction.

On an undistorted parent, the projected-shell theorem can be tested directly.
For \(q=5\), we compared a central \(5^3\) block and projected shells from
\(7^3\), \(9^3\), and \(11^3\) source blocks with the corresponding complete
overlapping construction.  Every shell added exactly 98 directions, and the
projectors onto the final 419-dimensional spaces agreed to
\(1.4\times10^{-15}\).

Next, we test how well the ideal span identity survives coordinate mapping.
As a typical test, we repeated the span comparison for the $q=5$ basis normally
used for a He atom.  In the ideal case, the overlap matrix of the combined
overlapping construction has many exactly zero eigenvalues.  Coordinate
distortion makes some of these eigenvalues small but nonzero, and so the
complete overlapping space becomes slightly larger.  Measured in the physical
overlap metric, the difference between the spaces remains small: the normalized
patch-union residual is $2.353\times10^{-3}$, and the least favorable retained
direction is rotated by only $0.334^\circ$ from the leading equal-dimensional
part of the union.  This small but measurable deviation is the completeness
cost of coordinate distortion.

Excellent completeness by itself does not ensure an accurate diagonal
interaction, which depends on the moment properties of each function.
Projection onto the parent shell and symmetric orthogonalization alter the
moments of the individual functions even without distortion.  We therefore
test those moments separately.  For a final function $g$, define
\begin{equation}
\langle f\rangle_p=
\frac{\int f(\mathbf r)g^p(\mathbf r)\,\dd\mathbf r}
     {\int g^p(\mathbf r)\,\dd\mathbf r},
\qquad p=1,2.
\label{eq:pqs-weighted-averages}
\end{equation}
The $p=1$ weight is signed, while $p=2$ is the usual $g^2$ weight.  With
$\mathbf c_p=\langle\mathbf r\rangle_p$, the mean Cartesian rms width and
relative center shift are
\begin{equation}
\sigma^2=\frac{1}{3}\left\langle|\mathbf r-\mathbf c_2|^2\right\rangle_2,
\qquad d_{12}=\frac{|\mathbf c_1-\mathbf c_2|}{\sigma}.
\label{eq:pqs-center-measure}
\end{equation}
For $\Delta r_a=r_a-c_{1a}$, with $a=x,y,z$, the mean pure and mixed
quadratic coefficients are
\begin{align}
t_2&=\frac{1}{6\sigma^2}\sum_a
\left|\langle(\Delta r_a)^2\rangle_1\right|,\nonumber\\
t_{11}&=\frac{1}{3\sigma^2}\sum_{a<b}
\left|\langle\Delta r_a\Delta r_b\rangle_1\right|.
\label{eq:pqs-moment-measures}
\end{align}
The factors include averaging over the three Cartesian components and, for
$t_2$, the usual factor of one half in a quadratic Taylor term.
Table~\ref{tab:pqs-moment-diagnostics} gives block means of these three
dimensionless measures.

\begin{table}[t]
\caption{\label{tab:pqs-moment-diagnostics}Mean absolute dimensionless
center and quadratic-moment diagnostics for the undistorted and standard
distorted projected-shell constructions.  For each noncentral row, the shell
label gives the size of the parent cube from which it is projected.}
\squeezetable
\begin{ruledtabular}
\begin{tabular}{lrrr}
shell & \(d_{12}\) & \(t_2\) & \(t_{11}\) \\
\hline
\multicolumn{4}{c}{undistorted} \\
central \(5^3\) & \(1.07\times10^{-14}\) & \(1.45\times10^{-14}\) & \(7.83\times10^{-16}\) \\
\(7^3\) parent  & \(7.46\times10^{-4}\) & \(6.01\times10^{-3}\) & \(5.47\times10^{-4}\) \\
\(9^3\) parent  & \(3.18\times10^{-3}\) & \(2.74\times10^{-2}\) & \(2.51\times10^{-3}\) \\
\(11^3\) parent & \(6.09\times10^{-3}\) & \(5.95\times10^{-2}\) & \(4.96\times10^{-3}\) \\
\hline
\multicolumn{4}{c}{standard distorted} \\
central \(5^3\) & \(6.25\times10^{-2}\) & \(6.68\times10^{-2}\) & \(2.58\times10^{-15}\) \\
\(7^3\) parent  & \(7.88\times10^{-2}\) & \(4.45\times10^{-2}\) & \(2.55\times10^{-4}\) \\
\(9^3\) parent  & \(6.53\times10^{-2}\) & \(3.77\times10^{-2}\) & \(1.64\times10^{-4}\) \\
\(11^3\) parent & \(5.00\times10^{-2}\) & \(2.93\times10^{-2}\) & \(1.24\times10^{-4}\) \\
\end{tabular}
\end{ruledtabular}
\end{table}

Because $\mathbf c_1$ is the signed center, the linear central moments vanish.
The normalized mismatch $d_{12}$ measures the remaining first-moment defect
relative to the $g^2$ center.  The analogous mismatch was found to track the
diagonal-interaction error well for radial gausslets~\cite{White2026RadialGausslets}.  The next
moments are the quadratic ones shown in the table.  For the uniform
construction, the two centers agree to numerical precision in the central
cube.  Shell projection introduces a small mismatch that grows outward,
reaching 0.0061 function widths in shell 11.  With the standard coordinate
distortion, the mean mismatch is larger, between 0.050 and 0.079 widths,
including in the unprojected central cube.  Coordinate mapping is therefore
the main source of the \(d_{12}\) center mismatch in the practical basis.  For the
distorted construction, $t_2$ remains below 0.067 and $t_{11}$ below
$2.6\times10^{-4}$.

The quadratic entries are coefficients in a width-scaled Taylor expansion.
For an orbital-product density that varies on a length scale $L$ larger than
the function width, their contribution is further reduced by approximately
$(\sigma/L)^2$.  The diagonal approximation is used here only for the electron--electron interaction,
where the orbital products entering the Coulomb integrals normally vary
smoothly over one gausslet.  Thus dimensionless moment defects of a few
$10^{-2}$ are small local corrections, not percent-level interaction-energy
errors.  The table therefore diagnoses the basis functions rather than
estimating an interaction or energy error; the molecular tests below measure the latter
directly.  The undistorted rows also show that exact equality of
the collective spans does not require every final function to retain exact
moments.

To test the combined effect of shell projection and coordinate mapping, we
studied how well the bases fit a family of Gaussian-type functions resembling
conventional basis functions.  We used \(q=5\) PQS and its equal-size
WL counterpart of order \(q-2=3\), both constructed from the same
mapped parent.  The family contains 220 functions: a Gaussian envelope of
width 3 bohr times all three-dimensional Chebyshev products through total
degree nine.  Chebyshev polynomials give better numerical scaling than
the usual monomial factors, with the same span.

Each test function is projected into and normalized within the common parent; its loss
is the fraction of squared norm discarded by contraction into the final basis.
This removes error already present in the parent.  Grouping the test functions by
total degree and number of active Cartesian directions gives 25 comparisons.
PQS has the smaller mean loss in all 25 groups, with a median reduction in
loss of more than a factor of four.  Thus the
higher local order of PQS preserves smooth orbital structure more accurately
at equal size even after projection and coordinate mapping.  No Hamiltonian
or interaction approximation enters this comparison; the molecular calculations below test whether
this advantage carries over to interactions and energies.

\section{Molecular tests of projected shells}
\label{sec:evidence}

Does the higher local order of PQS translate to more accurate electronic-structure calculations at fixed
basis size?  We compare PQS and WL in H$_2^+$, H$_2$, and
restricted C$_2$, progressing from a one-electron representation test to a
correlated two-electron problem and then a multi-orbital Hartree--Fock state.
We label each matched pair by \(n_s\): the full PQS contraction has
one-dimensional order \(q=n_s\), while the equal-size WL
construction has order \(n_s-2\) on its disjoint pieces.
Within each comparison the two bases use the same mapped parent, physical box,
reference state, and, where needed, Gaussian supplement.  All one-electron
matrix elements are evaluated in Galerkin form.  For the gausslet--gausslet
electron--electron interaction we use the integral diagonal approximation
(IDA) introduced with the original gausslets~\cite{White2017Gausslets}.  It
retains the diagonal form of Eq.~\eqref{eq:gausslet-diagonal}, but
defines its two-index matrix by integrating over the individual gausslets.
With \(w_i=\int G_i(\mathbf r)\,\dd\mathbf r\),
\begin{equation}
V_{ik}^{\rm IDA}=
\frac{1}{w_iw_k}\int\!\!\int
G_i(\mathbf r)\frac{G_k(\mathbf r')}{|\mathbf r-\mathbf r'|}
\,\dd\mathbf r\,\dd\mathbf r'.
\label{eq:ida-integral}
\end{equation}
Thus IDA averages the Coulomb kernel over the finite extent of both functions
rather than evaluating it only at their centers.  For compact near-nuclear
structure not resolved by the parent, we use the hybrid gausslet/Gaussian
construction of Qiu and White~\cite{QiuWhite2021}: atom-centered Gaussians are
projected into the space orthogonal to the final gausslet basis, and only the
linearly independent residual Gaussians are retained.  Terms involving
residual Gaussians use the matched-width Gaussian approximation introduced in
the same work.

\subsection{One-electron basis accuracy: H$_2^+$}
\label{subsec:h2plus}

With one electron, H$_2^+$ contains no electron--electron interaction, giving
a clean test of the information lost when a common parent space is replaced
by either nested basis.  One would normally add residual atom-centered
Gaussians, as in the H$_2$ calculations below.  We deliberately omit them
here so that the test isolates the two nesting constructions.  We use the
lowest states of even and odd
parity under inversion through the bond midpoint, denoted \(g\) and \(u\),
respectively.  The distorted parent has a
near-nuclear spacing of 0.30 bohr, an outer spacing of 2.8 bohr, and at least
10 bohr of padding.  At \(R=2\) bohr it contains 12789 functions.  Its \(g\)
and \(u\) energies lie 0.394 and 0.379 mHa above the corresponding
Riccati--Pad\'e references
\cite{FernandezGarcia2021H2plus,FernandezGarcia2021H2plusData}.

Every row of Table~\ref{tab:h2plus-r2} uses the same parent basis and
one-electron Hamiltonian, so the tabulated variational energy errors measure
the information discarded by the terminal bases separately from parent-basis
resolution.

\begin{table}[t]
\caption{\label{tab:h2plus-r2}One-electron H$_2^+$ energy errors at
\(R=2\) bohr relative to the common 12789-function parent basis.  Here
\(\Delta E_x=E_x^{\rm terminal}-E_x^{\rm parent}\); \(g\) and \(u\) denote
the lowest even-parity (gerade, bonding) and odd-parity (ungerade,
antibonding) states.}
\begin{ruledtabular}
\begin{tabular}{clrrr}
\(n_s\) & basis & \(N\) & \(\Delta E_g\) & \(\Delta E_u\) \\
 & & & (mHa) & (mHa) \\
\hline
4 & PQS & 923 & 4.6243 & 17.6105 \\
4 & WL & 923 & 31.8493 & 101.5540 \\
\hline
5 & PQS & 1285 & 0.2683 & 2.1703 \\
5 & WL & 1285 & 0.3652 & 8.0184 \\
\hline
6 & PQS & 1999 & 0.0412 & 0.2443 \\
6 & WL & 1999 & 0.2135 & 1.5049 \\
\end{tabular}
\end{ruledtabular}
\end{table}

Both constructions improve rapidly with order, but PQS is more accurate for
both states throughout.  For the compact \(n_s=4\) bases, it reduces the
\(g\) and \(u\) errors by factors of 6.9 and 5.8.  The more diffuse odd-parity
state is the sharper test: its error falls from 2.17 to 0.244 mHa between
\(n_s=5\) and 6 with PQS, compared with 8.02 to 1.505 mHa for
WL\@.

As a robustness check, we repeated the \(n_s=5\) comparison at core spacings
\(d=0.24\), 0.30, and 0.40 bohr.  For the standard nuclear map used here,
\(s=\sqrt{Zd}\), where \(Z\) is the nuclear charge; the corresponding
hydrogen mapping strengths are \(s=0.490\),
\(0.548\), and \(0.632\).  Although the errors vary nonmonotonically with spacing, PQS gives
the smaller energy error for both symmetries in all six
comparisons.

\subsection{Restricted Hartree--Fock comparison in H$_2$}
\label{subsec:h2interaction}

H$_2$ is the simplest molecule in which the electron--electron interaction
enters, giving a clean test of the diagonal interaction as well as the orbital
basis.  We consider 13 bond lengths from 1 to 8 bohr and orders
\(n_s=4,5,6\).  At each geometry and order, PQS and WL have equal
dimensions and are constructed from the same parent lattice.  Both are
supplemented by the contracted Cartesian \(s\) and \(p\) functions of
H/cc-pV6Z on the two atoms~\cite{Dunning1989,PetersonWoonDunning1994,Pritchard2019BSE}.  We follow the symmetric
restricted branch throughout; at stretched bonds this is a controlled RHF
comparison rather than a description of physical dissociation.

The RHF energy has two distinct sources of error: the finite basis changes the
orbital, and the IDA changes the Coulomb energy assigned to it.  We partially
separate these two effects for the two different forms of nesting in the following way: 
first, we perform the RHF optimization
for each type of nested-basis IDA Hamiltonian, getting a single occupied orbital for each.
Then we compare the interaction energy from that Hamiltonian with the interaction of the same
orbital projected into the parent basis. This estimates the IDA error for that orbital. Second,
we compare the RHF energy of the IDA system with an external reference energy, taken from
conventional bases.

Somewhat surprisingly for H$_2$, cc-pV6Z was too inaccurate to serve as a good reference, and required
augmentation.
For example, at \(R=8\) bohr, cc-pV6Z remains 0.327 mHa above the corresponding augmented basis,
aug-cc-pV6Z~\cite{KendallDunningHarrison1992,PetersonWoonDunning1994,Pritchard2019BSE}.  As our reference energy we extrapolated Cartesian aug-cc-pV5Z and
aug-cc-pV6Z PySCF energies~\cite{Sun2018PySCF} using
\(E_L=E_\infty+A(L+1)e^{-9\sqrt L}\)~\cite{KartonMartin2006HF}.
Their 0.007--0.018 mHa difference provides a rough energy error scale for the reference.

\begin{figure}[t]
\centering
\includegraphics[width=\columnwidth]{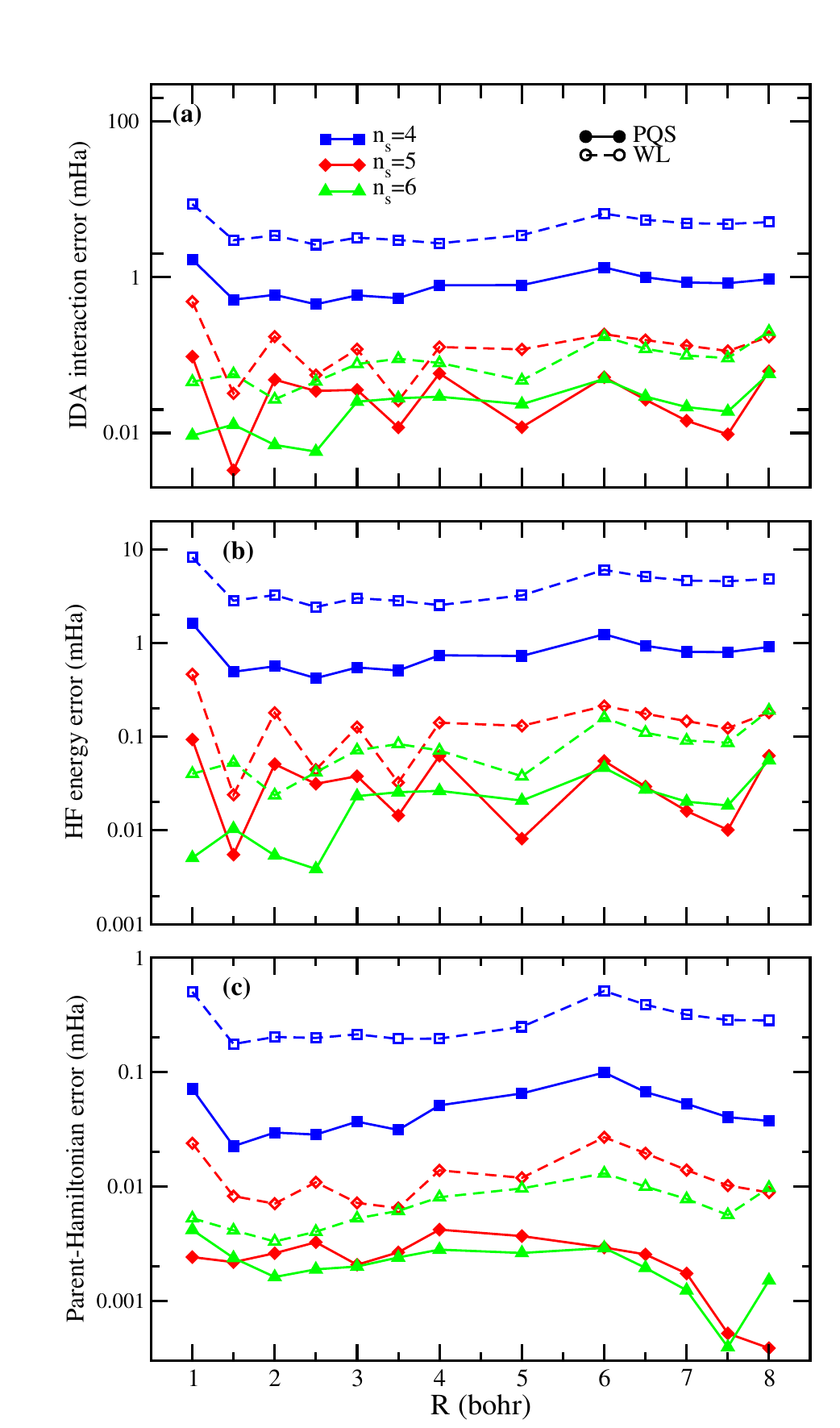}
\caption{\label{fig:h2-rhf-parent}Restricted-HF comparison for H$_2$
supplemented by contracted Cartesian H/cc-pV6Z \(s\) and \(p\) residual
Gaussians, at equal PQS and WL dimensions.  All three panels plot error
magnitudes.  Panel (a) compares the IDA Coulomb energy with the parent-basis
value for the same converged orbital.  Panel (b) uses the RHF energy obtained
with the IDA Hamiltonian; panel (c) reevaluates the same orbital with the
common parent Hamiltonian.  The reference for (b) and (c) is the HF basis-set
limit estimated from Cartesian aug-cc-pV5Z and aug-cc-pV6Z.  Solid filled
symbols denote PQS and dashed open symbols WL\@.}
\end{figure}

Figure~\ref{fig:h2-rhf-parent} shows the results.  Panel (a)
shows the IDA Coulomb error, panel (b) shows the error in the RHF energy relative to the reference.
Note that IDA energies can exhibit cancellation between basis incompleteness, which always raises the energy,
and IDA error, which can have either sign.  As a third measure of the errors, in panel (c)
we use the parent Hamiltonian to evaluate the energy of the orbital obtained with each method,
plotting the errors relative to the reference. For PQS, the errors at $n_s=5$ and $6$ match fairly well,
which probably means that the error of the reference is dominant.
PQS has the smaller IDA error and the lower parent-Hamiltonian orbital energy
at all 39 geometry/order points.  The negative WL IDA error
nevertheless makes its directly calculated RHF energy lower in 29 cases,
reversing the more meaningful orbital ordering.  Relative to the extrapolated
reference, PQS is closer in both panels (b) and (c) at all 39 points.
In short, PQS is clearly the more accurate form of nesting.

\subsection{Full-configuration-interaction comparison in H$_2$}
\label{subsec:h2correlated}

\begin{figure}[t]
\centering
\includegraphics[width=\columnwidth]{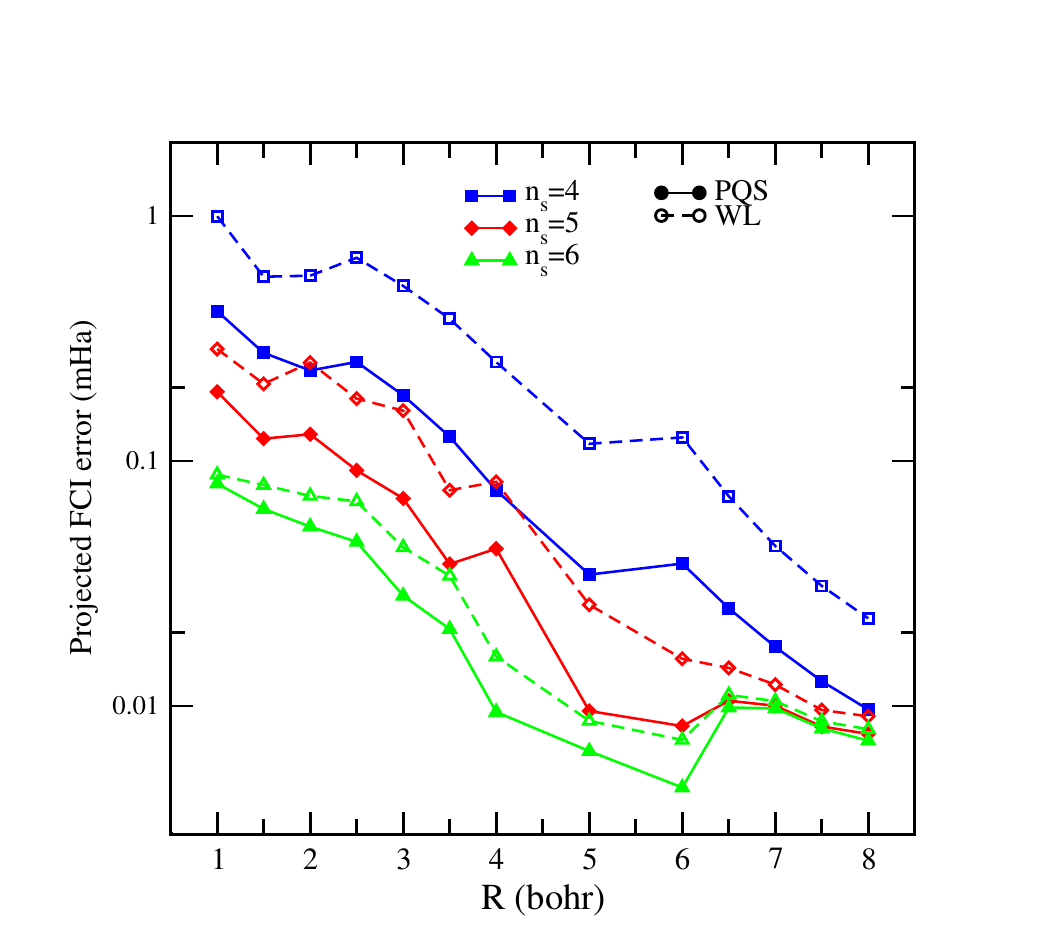}
\caption{\label{fig:h2-errors}Projected full-configuration-interaction (FCI)
errors for H$_2$ with the same H/cc-pV6Z \(s,p\) residual-Gaussian
supplement.  At each geometry, every lower-order FCI wave
function is projected into the same supplemented PQS \(n_s=7\) basis and its
energy is evaluated with the \(n_s=7\) Hamiltonian.  The plotted error is this
energy minus the FCI complete-basis-set (CBS) limit of Pachucki.  Colors and
symbols denote \(n_s\) as in
Fig.~\ref{fig:h2-rhf-parent}; solid filled symbols denote PQS and dashed open
symbols WL\@.  The source dimensions are equal at each geometry and
order.}
\end{figure}

The preceding test uses a single optimized determinant.  Figure~\ref{fig:h2-errors}
shows the corresponding full-configuration-interaction (FCI) comparison, in
which we solve the two-electron Hamiltonian exactly within each finite basis
using the same Gaussian supplement.  The errors obtained from each basis's own
Hamiltonian (not shown) have the same broad order dependence: PQS has a substantial
advantage at \(n_s=4\), a smaller one at \(n_s=5\), and the two constructions
are nearly indistinguishable at \(n_s=6\).  The individual curves are less
regular, however, because wave-function and interaction errors from different
Hamiltonians can cancel.  We therefore show in the figure the result obtained
by projecting every lower-order FCI state into the same supplemented PQS
\(n_s=7\) basis and evaluating it with the \(n_s=7\) Hamiltonian.  The plotted
quantity is the magnitude of this energy error
relative to the FCI complete-basis-set limit of Pachucki~\cite{Pachucki2010H2BO}.
The finite \(n_s=7\) Hamiltonian contributes a common offset at each geometry,
which affects the absolute errors but not the comparison between the two
constructions.

As above, PQS uses \(q=n_s\), whereas the WL side space contains
\(n_s-2\) functions, giving equal dimensions at each geometry and order.  The
parent lattice and Gaussian supplement are otherwise identical.

PQS gives the smaller projected FCI error at every plotted geometry and order.
Its advantage is largest at \(n_s=4\), remains substantial at \(n_s=5\), and becomes
small at \(n_s=6\).  The errors broadly decrease with bond length.  Although
stretched H$_2$ is strongly correlated, its electrons are increasingly found
on different atoms, reducing the short-range correlation structure that the
basis must resolve.

\subsection{Multi-orbital restricted-HF comparison: C$_2$}
\label{subsec:c2rhf}

The carbon dimer provides a more stringent one-determinant test than H$_2$.
Its restricted state has six occupied spatial orbitals, including core and
valence orbitals of different angular character, and substantial direct and
exchange contributions.  We consider two bond lengths and four orders.  At
each geometry and order, the two constructions use the same mapped parent,
with core spacing \(1.2/[6(n_s-1)]\) bohr, and the contracted Cartesian
\(s\) and \(p\) functions from C/cc-pV6Z on each atom
\cite{WilsonVanMourikDunning1996,Pritchard2019BSE}.  PQS has order \(q=n_s\) and
WL has \(q=n_s-2\).  Their final dimensions are equal except for small
differences caused by residual Gaussian functions that fall below the singular-value cutoff.

Table~\ref{tab:c2-rhf} gives the RHF energies obtained directly from each IDA
Hamiltonian.  At each geometry, the reference is the internally stable real
restricted-HF branch, extrapolated from the more compact spherical, rather than
Cartesian, aug-cc-pV5Z and aug-cc-pV6Z PySCF calculations
\cite{KendallDunningHarrison1992,Sun2018PySCF}, using the same extrapolation as
for H$_2$.  The estimated reference uncertainties are 0.03 and 0.02 mHa at the
two geometries; since the references are common to PQS and WL, these affect only
the individual signed errors, not their comparison.  The estimated remaining HF
convergence error is below 0.005 mHa for every row.  The restricted branches are
unstable to complex and unrestricted variations; this is therefore a controlled
restricted-state basis benchmark, not a ground-state claim.

\begin{table}[t]
\caption{\label{tab:c2-rhf}Signed restricted-HF energy errors for C$_2$,
in mHa, relative to the extrapolated aug-cc-pV5Z/aug-cc-pV6Z references.
Positive values lie above the reference.  \(N_{\rm P}/N_{\rm WL}\) gives the
final PQS and WL dimensions.}
\begin{ruledtabular}
\begin{tabular}{ccrrr}
\(R\) & \(n_s\) & \(N_{\rm P}/N_{\rm WL}\) & \(\Delta E_{\rm P}\) & \(\Delta E_{\rm WL}\) \\
(bohr) & & & (mHa) & (mHa) \\
\hline
2.35 & 5 & 2119/2119 & -4.470 & -33.488 \\
2.35 & 6 & 3684/3684 & -3.377 & -12.086 \\
2.35 & 7 & 5427/5433 & +0.164 &  -0.073 \\
2.35 & 8 & 8056/8060 & -0.059 &  -0.227 \\
\hline
3.00 & 5 & 2249/2249 & -4.478 & -26.814 \\
3.00 & 6 & 3770/3770 & -5.195 & -17.336 \\
3.00 & 7 & 5523/5529 & +0.192 &  -0.320 \\
3.00 & 8 & 8262/8266 & -0.110 &  -0.315 \\
\end{tabular}
\end{ruledtabular}
\end{table}

The advantage of the larger $q$ from PQS is largest for small $n_s$.
At \(n_s=5\), the WL error magnitude is about 7.5 and 6 times the PQS value at
\(R=2.35\) and 3.00 bohr, respectively; PQS remains better at \(n_s=6\).  By \(n_s=7\),
both constructions have entered the sub-mHa regime.  Their signed errors then
change nonmonotonically because finite-basis and diagonal-interaction errors
can cancel, and the small error at $R=2.35$, $n_s=7$ for WL is likely due to cancellation.
From the results overall, it is clear that PQS is once again superior to WL\@.

\FloatBarrier
\section{Conclusions}
\label{sec:conclusions}

Dividing WL shells into disjoint pieces lowers both polynomial completeness and
the moment order from \(q\) to \(q-2\).  Here we have shown how
projected \(q\)-shells approximately restore the two lost orders and provide superior
results for several test systems.  The foundation of the PQS construction is a theorem showing that
in the ideal undistorted case, the span of PQS matches exactly that of a set of overlapping
patches of functions, where the completeness is intuitively apparent.  Besides completeness, the
basis functions need a set of low-order moments to vanish for an accurate diagonal approximation
to the interaction.  Although the theorem does
not protect the moment properties under the projection and symmetric orthogonalization used in PQS, we
give an argument explaining why the resulting moment errors should remain small,
keeping errors in the diagonal approximation small.
Our span and moment tests verify that these departures from the
ideal are small compared with the gain from retaining the full
local order.

The electronic-structure comparisons test progressively more of the method.
H$_2^+$ isolates reduction of a common parent basis, H$_2$ adds the approximate
interaction and a correlated molecular curve, and restricted C$_2$ adds core
and valence orbitals together with substantial exchange.  At matched or
nearly matched dimension, PQS gives a clear overall accuracy advantage in all
three settings.  Its advantage is largest in compact bases and decreases as
both constructions approach the same high-order limit.

PQS reorganizes information already present in the parent lattice; it cannot
recover short-distance structure that the parent never represented.  Residual
Gaussian functions therefore remain an independent and complementary part of
the molecular basis.  Interaction errors must likewise be distinguished from
one-particle errors.  The parent-Hamiltonian evaluations in H$_2$ make this
distinction explicit: the larger negative interaction error of
WL can reverse the ordering of the optimized orbitals.  In C$_2$,
cancellation between finite-basis and interaction errors makes the high-order
signed energies nonmonotonic, while the full two-geometry ladder shows the
large advantage of PQS for compact bases.  Taken together, the theorem,
construction tests, and molecular calculations show that PQS is a superior construction
at essentially no dimensional cost.

\appendix
\section{Proof of the projected-shell theorem}
\label{app:patchproof}

Pictorially, the projected-shell theorem says that, in the ideal construction
considered here, the functions shown in
Fig.~\ref{fig:multiresolution-patches}(c) and those retained in
Fig.~\ref{fig:multiresolution-patches}(d) span exactly the same space.
Figure~\ref{fig:multiresolution-patches} shows the construction in one
dimension.  In higher dimensions, the complete lines of functions in
Fig.~\ref{fig:multiresolution-patches}(c) become filled product patches---squares
in two dimensions and cubes in three---while the two retained functions at
each scale in Fig.~\ref{fig:multiresolution-patches}(d) become boundary shells,
as illustrated in Fig.~\ref{fig:pqs-2d-shell}.  The functions are COMX
contractions of finite regions of a uniform parent lattice of ideal gausslets.
We define these objects and prove the result below.

\subsection{A segment of an ideal gausslet chain}

We first work in one dimension.  Imagine an infinite, unit-spaced chain
of ideal gausslets
\begin{equation}
\ldots,g_{-2}(x),g_{-1}(x),g_0(x),g_1(x),g_2(x),\ldots .
\end{equation}
Gausslet \(g_i\) is centered at the integer \(i\).
Fix an integer order \(q\ge3\).  In the ideal model, the following relations
hold exactly; the last is required for every polynomial \(p\) of degree at
most \(q-1\):
\begin{equation}
 \begin{aligned}
 \langle g_i|g_j\rangle&=\delta_{ij},\\
 \langle g_i|\hat x|g_j\rangle&=i\,\delta_{ij},\\
 \langle g_i|p\rangle&=p(i),\qquad \deg p\le q-1.
 \end{aligned}
 \label{eq:ideal-gausslet-coordinate-matrix}
\end{equation}
Here \(\hat x\) is the ordinary continuum position operator, and
\(\langle g_i|p\rangle=\int g_i(x)p(x)\,\dd x\).  Thus the coefficient of a
polynomial on gausslet \(g_i\) is its value at the gausslet center.  Given an
ordinary polynomial \(p(x)\), define its representation on a finite segment by
\begin{equation}
 \Phi_R[p](x)=\sum_{i=-R}^{R}p(i)g_i(x).
 \label{eq:ideal-patch-function}
\end{equation}
Here \(R\) is a positive integer, and \(p(i)\) means simply that the polynomial
is evaluated at \(x=i\).

From these parent gausslets we form the order-\(q\) patch space
\begin{equation}
 K_R^{(q)}=\Span\left\{
 \Phi_R[1],\Phi_R[x],\ldots,\Phi_R[x^{q-1}]
 \right\}.
 \label{eq:ideal-local-space}
\end{equation}
Thus \(K_R^{(q)}\) consists of \(q\) polynomially weighted linear combinations
of the \(2R+1\) parent gausslets.  These combinations are linearly independent
provided \(2R+1\ge q\): a polynomial of degree at most \(q-1\) that vanishes
at \(q\) or more distinct sites must be identically zero.  Hence
\(\dim K_R^{(q)}=q\).  We will compare this space with the patch on the next
inner segment, from \(-R+1\) through \(R-1\).  The condition
\begin{equation}
 2R-1\ge q
 \label{eq:inner-site-condition}
\end{equation}
just says that this smaller segment still contains enough gausslets to carry a
complete \(q\)-dimensional polynomial patch.

\subsection{COMX functions as a polynomial discrete-variable representation}

We next establish explicitly the relation between COMX functions and cardinal
polynomials.  Orthonormality of the ideal gausslets gives
\begin{equation}
 \left\langle\Phi_R[p],\Phi_R[r]\right\rangle
 =\sum_{i=-R}^{R}p(i)r(i).
 \label{eq:discrete-polynomial-inner-product}
\end{equation}
The right-hand side defines a positive-definite inner product for real
polynomials of degree at most \(q\).  A nonzero polynomial in this space
cannot have zero norm, because the segment contains more than \(q\) distinct
sites.

Let \(\pi_q(x)\) be the degree-\(q\) polynomial obtained by orthogonalizing
\(x^q\) against \(1,x,\ldots,x^{q-1}\).  It obeys
\begin{equation}
 \sum_{i=-R}^{R}\pi_q(i)s(i)=0,
 \qquad \deg s<q.
 \label{eq:next-orthogonal-polynomial}
\end{equation}
For completeness, we show that \(\pi_q\) has exactly \(q\) distinct real roots.
Suppose instead that it has only \(m<q\) distinct real roots at which it changes
sign, i.e., roots of odd multiplicity, and call them
\(\xi_1,\ldots,\xi_m\).  The polynomial
\begin{equation}
 s_m(x)=\prod_{j=1}^{m}(x-\xi_j)
 \label{eq:sign-change-polynomial}
\end{equation}
has degree less than \(q\).  Thus Eq.~\eqref{eq:next-orthogonal-polynomial}
gives
\[
 \sum_{i=-R}^{R}s_m(i)\pi_q(i)=0.
\]
But \(s_m\pi_q\) cannot change sign.  Its distinct real zeros are among those
of the degree-\(q\) polynomial \(\pi_q\), so there are at most \(q\) of them,
whereas the segment contains \(2R+1>q\) parent sites.  Thus the product cannot
vanish at every parent site, and the sum cannot be zero.  Hence \(m\ge q\).
Since
\(\pi_q\) has degree \(q\), it must have exactly \(q\) sign-changing roots.
They are consequently real, distinct, and simple.  Order them as
\begin{equation}
 \lambda_1<\lambda_2<\cdots<\lambda_q.
 \label{eq:ordered-comx-centers}
\end{equation}

For each root, define the degree-\((q-1)\) cardinal polynomial
\begin{equation}
 \ell_a(x)=
 \prod_{\substack{b=1\\b\ne a}}^{q}
 \frac{x-\lambda_b}{\lambda_a-\lambda_b}.
 \label{eq:cardinal-polynomials}
\end{equation}
It obeys \(\ell_a(\lambda_b)=\delta_{ab}\).  Since
\(\pi_q(x)=\prod_{b=1}^q(x-\lambda_b)\),
Eq.~\eqref{eq:cardinal-polynomials} implies
\begin{equation}
 (x-\lambda_a)\ell_a(x)
 =\frac{\pi_q(x)}{\pi_q'(\lambda_a)}.
 \label{eq:cardinal-orthogonal-identity}
\end{equation}
Here \(\pi_q'(\lambda_a)\) is the derivative of \(\pi_q(x)\) evaluated at
\(x=\lambda_a\).

Since \(\ell_b\) has degree \(q-1\),
Eqs.~\eqref{eq:discrete-polynomial-inner-product} and
\eqref{eq:next-orthogonal-polynomial} give
\[
 \left\langle\Phi_R[\ell_b]\middle|\Phi_R[\pi_q]\right\rangle=0.
\]
Applying \(\Phi_R\) to Eq.~\eqref{eq:cardinal-orthogonal-identity}, taking
its inner product with \(\Phi_R[\ell_b]\), and using
Eq.~\eqref{eq:ideal-gausslet-coordinate-matrix} then gives
\begin{equation}
 \left\langle\Phi_R[\ell_b]\middle|\hat x\middle|\Phi_R[\ell_a]\right\rangle
 =\lambda_a
 \left\langle\Phi_R[\ell_b]\middle|\Phi_R[\ell_a]\right\rangle.
 \label{eq:cardinal-is-comx}
\end{equation}
Hermiticity gives the same left-hand side with \(\lambda_b\) in place of
\(\lambda_a\), and hence
\[
 (\lambda_a-\lambda_b)
 \left\langle\Phi_R[\ell_b]\middle|\Phi_R[\ell_a]\right\rangle=0.
\]
The distinct \(\lambda_a\) therefore make the \(q\) functions
\(\Phi_R[\ell_a]\) mutually orthogonal, while
Eq.~\eqref{eq:cardinal-is-comx} shows that the coordinate matrix is diagonal
in this basis, with centers \(\lambda_a\).  These are the COMX functions, up
to arbitrary individual scale factors.

Taking \(a=b\) in Eq.~\eqref{eq:cardinal-is-comx} and dividing by the norm
squared of \(\Phi_R[\ell_a]\) shows that \(\lambda_a\) is the coordinate
expectation value of the corresponding normalized COMX function.  Since the
parent centers run from \(-R\) to \(R\), each \(\lambda_a\) lies in that interval.
Equality at either endpoint would require the function to be \(g_{-R}\) or
\(g_R\).  Neither belongs to \(K_R^{(q)}\), since its coefficient pattern
would require a polynomial of degree at most \(q-1\) to vanish at the other
\(2R\) sites.  Thus all the \(\lambda_a\) lie strictly between \(-R\) and
\(R\).  This is the standard polynomial discrete-variable-representation
construction, here applied to the discrete measure on the ideal gausslet
sites.

We call the normalized functions associated with \(\ell_1\) and \(\ell_q\)
the left and right edge functions, \(e_L\) and \(e_R\).

\subsection{What an outer patch adds}

\begin{figure}[t]
\centering
\includegraphics[width=0.98\columnwidth]{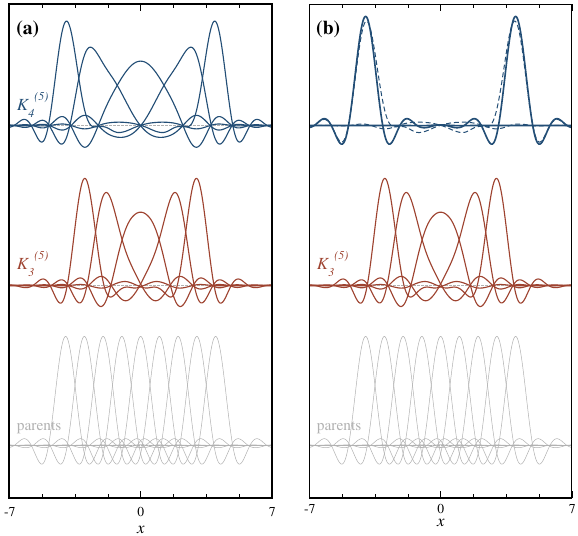}
\caption{\label{fig:one-dimensional-projected-shell}Numerical
realization for an undistorted, unit-spaced parent chain, an outer segment with
\(R=4\), and \(q=5\).  The nine parent gausslets are gray.  (a) The five
outer-patch COMX functions (blue) and five inner-patch functions (red).
(b) The original outer edge functions (dashed) and their projections onto the
two endpoint parent gausslets forming the parent-lattice shell (solid),
together with the inner-patch functions.  The complete-patch union and the
projected construction both have rank seven, and their projector residual is
\(1.12\times10^{-15}\), illustrating Eq.~\eqref{eq:patch-shell-span}.
Vertical offsets are for clarity.}
\end{figure}

Figure~\ref{fig:one-dimensional-projected-shell} shows the basic result in one
dimension.  The outer and inner patches are contractions of the same parent
chain.  Projecting the two edge COMX functions \(e_L\) and \(e_R\) onto the
parent-lattice shell keeps only their coefficients on the two endpoint
gausslets.  Those projected functions span the two endpoint gausslets, so the
outer patch adds exactly two directions to the inner one.

The same result in any number of dimensions follows most directly by splitting
the sum over parent sites.  Let the Cartesian axes be labeled by \(\alpha\),
with order \(q_\alpha\ge3\) on each axis.  A cubic construction has
\(q_\alpha=q\), while an elongated molecular block has orders \(q,q,L\).
For a parent site \(\mathbf i=(i_1,\ldots,i_d)\), define the product gausslet
\begin{equation}
 g_{\mathbf i}(\mathbf x)=\prod_{\alpha=1}^{d}g_{i_\alpha}(x_\alpha).
 \label{eq:product-gausslet}
\end{equation}
Let \(B_{\mathbf R}\) be the rectangular block with
\(-R_\alpha\le i_\alpha\le R_\alpha\) on every axis.  Its inner block,
\(B_{\mathbf R-\mathbf 1}\), has every half-width reduced by one.  The
\emph{parent shell} is the one-site-thick layer of sites in the outer block
that are not in the inner block.  Thus a shell site has
\(|i_\alpha|=R_\alpha\) on at least one axis.

Let \(\mathcal P_{\mathbf q}\) be the space of polynomials having degree at
most \(q_\alpha-1\) in coordinate \(x_\alpha\), separately on every axis.
It contains \(\prod_\alpha q_\alpha\) independent polynomials.  The product
patch generated by a polynomial \(p\) is
\begin{align}
 \Phi_{\mathbf R}[p]
 &=\sum_{\mathbf i\,\mathrm{on}\,B_{\mathbf R}}
 p(\mathbf i)g_{\mathbf i},\notag\\
 K_{\mathbf R}
 &=\Span\bigl(\Phi_{\mathbf R}[p]
 \text{ for all }p\text{ in }\mathcal P_{\mathbf q}\bigr).
 \label{eq:product-patch-map}
\end{align}
We assume
\begin{equation}
 2R_\alpha-1\ge q_\alpha
 \qquad\text{on every axis},
 \label{eq:product-inner-site-condition}
\end{equation}
so the inner block has enough sites to carry the full polynomial patch.  A
polynomial having degree at most \(q_\alpha-1\) in \(x_\alpha\) for every axis
\(\alpha\) that vanishes on the inner grid must be zero.  This follows by
applying one-dimensional polynomial interpolation successively along each
coordinate.  Distinct polynomials therefore give distinct patch functions in
both the outer and inner blocks, and both patch spaces have dimension
\(\prod_\alpha q_\alpha\).

Let \(\mathcal Q_{\mathbf R}\) keep the coefficients on the parent shell and
set all inner-block coefficients to zero.  In the ideal model this is an
orthogonal projection because the parent gausslets are orthonormal.  The
rectangular matrix \(Q\) in Eq.~\eqref{eq:lowdin} is the same coefficient
restriction with its zero rows omitted.  Splitting the site sum in
Eq.~\eqref{eq:product-patch-map} gives, for every \(p\),
\begin{equation}
 \Phi_{\mathbf R}[p]
 =\Phi_{\mathbf R-\mathbf 1}[p]
  +\mathcal Q_{\mathbf R}\Phi_{\mathbf R}[p].
 \label{eq:patch-shell-splitting}
\end{equation}
The first term is the same polynomial restricted to the inner parent sites;
it is not an orthogonal projection onto the inner patch space.  The second
term is supported only on the parent shell.  Equation~\eqref{eq:patch-shell-splitting}
therefore gives
\begin{equation}
 \Span(K_{\mathbf R},K_{\mathbf R-\mathbf 1})
 =\Span(K_{\mathbf R-\mathbf 1},
        \mathcal Q_{\mathbf R}K_{\mathbf R}).
 \label{eq:patch-shell-span}
\end{equation}
The two spaces on the right have disjoint parent support.  They therefore
share no nonzero direction, so no direction is counted twice.

It remains to count the directions in the projected parent-shell space.
The shell part of \(\Phi_{\mathbf R}[p]\) vanishes exactly when \(p\) vanishes
on every shell site.  Consider one face, such as
\(i_\alpha=R_\alpha\).  On that face, \(p\) vanishes on a grid containing at
least \(q_\beta\) sites along every other axis.  It must therefore vanish as
a polynomial on the whole face.  The same reasoning applies to both faces on
every axis.  Treating \(p\) as a polynomial in \(x_\alpha\), the ordinary
factor theorem then supplies the factors \(x_\alpha-R_\alpha\) and
\(x_\alpha+R_\alpha\).  Repeating this for every axis gives
\begin{equation}
 p(\mathbf x)=
 \left[\prod_{\alpha=1}^{d}(x_\alpha^2-R_\alpha^2)\right]
 s(\mathbf x),
 \label{eq:shell-vanishing-polynomial}
\end{equation}
where \(s\) has degree at most \(q_\alpha-3\) in \(x_\alpha\).  Conversely,
every polynomial of this form vanishes on the parent shell.  These polynomials
form a space of dimension \(\prod_\alpha(q_\alpha-2)\).  Subtracting this from
the full patch dimension gives
\begin{equation}
 \dim\mathcal Q_{\mathbf R}K_{\mathbf R}
 =\prod_{\alpha=1}^{d}q_\alpha
  -\prod_{\alpha=1}^{d}(q_\alpha-2).
 \label{eq:projected-shell-dimension}
\end{equation}

In one dimension this gives two new directions, and the combined inner and
outer patches have dimension \(q+2\), as in
Fig.~\ref{fig:one-dimensional-projected-shell}.  In two dimensions the count
is \(q^2-(q-2)^2\).  In three dimensions it is
\(q^3-(q-2)^3\); for \(q=5\), the 98 functions may be grouped as 54 face,
36 edge, and 8 corner functions.  For a molecular block with orders
\(q,q,L\), the same equation gives
\begin{equation}
 q^2L-(q-2)^2(L-2),
 \label{eq:rectangular-shell-count-proof}
\end{equation}
which is Eq.~\eqref{eq:pqsdim}.

\subsection{Why the selected COMX shell is sufficient}

Equation~\eqref{eq:projected-shell-dimension} counted the projection of the
whole patch.  PQS projects only the COMX products having an outer COMX index
on at least one axis.  We now show that this selected COMX shell supplies all
of the directions counted above.

On axis \(\alpha\), denote the COMX nodes by
\(\lambda_{\alpha 1}<\cdots<\lambda_{\alpha q_\alpha}\) and their cardinal
polynomials by \(\ell_{\alpha b}\).  The results above show that all of these
nodes lie strictly between \(-R_\alpha\) and \(R_\alpha\).  The product
cardinal expansion of any \(p\) in \(\mathcal P_{\mathbf q}\) is
\begin{equation}
 p(\mathbf x)=
 \sum_{b_1=1}^{q_1}\!\cdots\!\sum_{b_d=1}^{q_d}
 p(\lambda_{1b_1},\ldots,\lambda_{db_d})
 \prod_{\alpha=1}^{d}\ell_{\alpha b_\alpha}(x_\alpha).
 \label{eq:product-cardinal-expansion}
\end{equation}
The COMX shell consists of the products in this sum for which at least one
index \(b_\alpha\) equals 1 or \(q_\alpha\).  Its size is exactly the
right-hand side of Eq.~\eqref{eq:projected-shell-dimension}.

Suppose a combination of these selected products vanished after restriction
to the parent shell.  Its polynomial \(p\) would then have the factorized form
in Eq.~\eqref{eq:shell-vanishing-polynomial}.  The coefficients of the
all-interior cardinal products are absent; equivalently, \(p\) vanishes at
every COMX-node tuple made from the interior nodes
\(\lambda_{\alpha 2},\ldots,\lambda_{\alpha,q_\alpha-1}\).  Each factor
\(x_\alpha^2-R_\alpha^2\) is nonzero at every interior node, because all such
nodes lie strictly between \(-R_\alpha\) and \(R_\alpha\).  Hence \(s\)
vanishes on a grid of \(q_\alpha-2\) distinct points on every axis.  Applying
the same one-dimensional zero-counting argument successively along every axis
shows that \(s\) must be zero: in coordinate \(x_\alpha\) its degree is at
most \(q_\alpha-3\), while the grid supplies \(q_\alpha-2\) distinct nodes.
Therefore \(p\) is also zero.  Thus
no nonzero combination of selected COMX-shell functions disappears under the
parent-shell projection.  Since their number equals the dimension in
Eq.~\eqref{eq:projected-shell-dimension}, their projections span the full
projected-patch shell space \(\mathcal Q_{\mathbf R}K_{\mathbf R}\).  The
symmetric orthogonalization in
Eq.~\eqref{eq:lowdin} changes the functions but not their span.

\subsection{Repeated nesting and practical qualifications}

Take an innermost complete product patch and surround it by successively
larger complete patches made from the same ideal parent gausslets.  Denote the
patches, from inside out, by
\(\mathcal B_j=K_{\mathbf R_j}\), with
\(\mathbf R_j=\mathbf R_0+j\mathbf 1\), where \(\mathbf 1\) adds one
parent-site layer on every axis.  Let \(S_j\) be the space spanned by the
selected COMX-shell functions of \(\mathcal B_j\), and let
\(\Sigma_j=\mathcal Q_{\mathbf R_j}S_j\) be its projection onto the
corresponding parent shell.  The preceding result shows that
\(\Sigma_j=\mathcal Q_{\mathbf R_j}K_{\mathbf R_j}\) as spaces.  Applying
Eq.~\eqref{eq:patch-shell-span} at each scale gives the projected-shell
theorem,
\begin{equation}
 \Span(\mathcal B_0,\mathcal B_1,\ldots,\mathcal B_M)
 =\Span(\mathcal B_0,\Sigma_1,\ldots,\Sigma_M).
 \label{eq:full-patch-nesting}
\end{equation}
The right-hand side keeps the innermost complete patch and only the new shell
directions introduced at each larger scale.  The theorem concerns the span of
the chosen patch spaces; it does not claim completeness in the full parent
lattice or in the continuum.  After shell projection, each parent-lattice
site occurs in only one retained shell.

Actual gausslets differ from the ideal model in two relevant ways.  First,
their polynomial completeness, orthogonality, and projected-coordinate
properties are numerical rather than exact.  On a uniform parent lattice the
ideal rank and span relations are nevertheless reproduced to numerical
precision, as shown in Sec.~\ref{subsec:pqs-construction-tests}.  This test
supplies the direct bridge between the ideal functions used in the proof and
the gausslets used in the calculation.

Second, the molecular construction applies a nonlinear coordinate map to the
uniform parent chain.  A polynomial in the physical coordinate then no longer
has an exactly low-degree coefficient envelope in the uniform site label.
The pieces removed from an outer patch are consequently not represented
exactly by the next inner patch, and the complete overlapping-patch union can
be slightly larger than the PQS space.  The standard-map calculation in
Sec.~\ref{subsec:pqs-construction-tests} measures this departure directly:
the 419-function PQS space has a normalized patch-union residual of
\(2.353\times10^{-3}\) and a maximum principal angle of \(0.334^\circ\) from
the leading union subspace.

Finally, equality of spans says nothing by itself about the moment properties
of each localized basis function.  Those properties are needed for the
diagonal interaction approximation and are tested separately in
Table~\ref{tab:pqs-moment-diagnostics}.  The projected-shell theorem explains
the retained local order; the moment and smooth-function tests quantify the
two approximations made when that ideal construction is turned into the
practical PQS molecular basis.

The basis construction and Hamiltonian generation were performed with the open
source \texttt{GaussletBases.jl} software library at
\url{https://github.com/srwhite59/GaussletBases.jl}.

\acknowledgments
I thank Sandeep Sharma for helpful discussions. This work was supported by the
U.S. NSF under Grant DMR-2412638.

\section*{AUTHOR DECLARATIONS}

\subsection*{Conflict of Interest}
The author has no conflicts to disclose.

\subsection*{Author Contributions}
\textit{Steven R. White}: Conceptualization; Methodology; Software;
Validation; Formal analysis; Investigation; Data curation; Visualization;
Writing---original draft; Writing---review and editing; Funding acquisition.

\section*{DATA AVAILABILITY}
The data supporting this study are available from the author upon reasonable
request.  The open-source PQS implementation is available in
\texttt{GaussletBases.jl} at
\url{https://github.com/srwhite59/GaussletBases.jl}.

\bibliographystyle{aipnum4-2}
\bibliography{ref}

\end{document}